# Quantitative trends in 8 physical properties of 115000 inorganic compounds gained by machine learning

Evgeny Blokhin [1] and Pierre Villars [2]

1 Tilde Materials Informatics, Straßmannstraße 25, 10249, Berlin, Germany, and
Materials Platform for Data Science, Sepapaja 6, 15551, Tallinn, Estonia
2 Material Phases Data System, Unterschwanden 6, 6354, Vitznau, Switzerland

Emails: eb@tilde.pro and villars.mpds@bluewin.ch

## Abstract

We applied the decision trees (random forest) machine-learning technique for the large experimental materials dataset PAULING FILE, compiled from the world's peer-reviewed literature. The training and validation data were extracted from the hundreds of thousands of publications in materials science (1891―2017). Then, for the nearly 115'000 distinct inorganic compounds we predicted 8 thermodynamic, mechanical, and electronic properties, using the only crystalline structures as an input. For the predicted physical properties we observed certain periodical patterns in all unary, binary, ternary, and quaternary compounds. We also solved a reversed task of predicting the possible crystalline structure based on a given combination of values of the 8 mentioned properties. Therefore our observations may play a role of the periodic table, formulated not for the chemical elements, but for the entire set of materials.

## Introduction

Analogous to the Human Genome Project in 90-s, today's efforts in materials informatics [1-5] tackle the "Materials Genome", producing the vast swatches of data to be validated and interpreted. In view of this we emphasize the following critical statement. For any predictive data-intensive technique, the combination of *quality* and *quantity* of the input data must lead to the results comparable with the first-principles quantum-mechanics predictions. Indeed, it is known that the input quality ("garbage in — garbage out"), as well as the input quantity ("weak algorithms with more data beat better algorithms with less data") are crucial [6, 7]. Herewith we propose a pathway to decipher the "Materials Genome" using a relatively cheap and unsophisticated machine learning technique – decision trees – and the large materials dataset of an exceptional quality. We exemplify the predictions of the 8 physical properties for the nearly 115'000 inorganic compounds from the only crystalline structures, as well as the reverse predictions of the crystalline structures from the values of these 8 physical properties. The considered 8 physical properties are: (*a*) isothermal bulk modulus, (*b*) enthalpy of formation, (*c*) heat capacity at constant pressure, (*d*) Seebeck coefficient, (*e*) temperature for congruent melting, (*f*) Debye temperature, (*g*) linear thermal expansion coefficient, and (*h*) energy gap (direct or indirect) for insulators.

Currently, the materials informatics is a collection of recipes taken from the computer science and adopted for the materials science. The main difficulty is purely technical from an academic point of view — how to efficiently handle materials big data. A series of modern initiatives in materials informatics is known, both academic and industrial [1-5]. They all have one main feature in common: they build their own software infrastructures to combat the challenge of materials big data, thus gaining the insight knowledge more efficiently. Probably the oldest effort is the PAULING FILE project [8], which was launched in 1993 as a joint venture of the Japan Science and Technology Corporation, Material Phases Data System (Switzerland), and The University of Tokyo, RACE.

Three goals were selected by the PAULING FILE project from the very beginning. The first goal was to create and maintain a comprehensive critically evaluated database for inorganic crystalline substances, covering crystallographic data, diffraction patterns, intrinsic physical properties and phase diagrams. The data had to be manually extracted from the peer-reviewed articles, books, proceedings *etc.* and additionally checked with extreme care. The term "inorganic substances" was defined as the compounds without C-H bonds. In parallel to the database creation, the second goal was to develop an appropriate retrieval software to make the data accessible in a convenient graphical user interface. In the longer term, as the third goal, the new tools for materials design should be created, *e.g.* to automatically search the database for correlations. Currently, all these goals are very close to being fulfilled. During the past 25 years, almost 300'000 publications (1891—2017) were manually processed, and about 400'000 crystalline structures, 60'000 phase diagrams, and 1'000'000 other physical property sets were extracted. The parts of the PAULING FILE data are included in many commercial products, such as Springer Materials and MedeA Materials Design. Today the PAULING FILE project is quite well-known, and there are already an order of thousand of publications referring it. The recent implementation of the PAULING FILE retrieval software and the materials design tools is an online product titled Materials Platform for Data Science (MPDS) [9].

The PAULING FILE database has the following structure. The standard unit of data is called an *entry*. All the entries are subdivided into three kinds: crystalline structures, physical properties, and phase diagrams. Another dimension of the data is the *distinct phases*. The three kinds of entries are interlinked via the distinct phases to which they belong. A tremendous amount of work was conducted by the PAULING FILE team during the past 25 years to manually distinguish about 140'000 distinct phases of inorganic materials, appearing in the world's scientific literature since 1891. Each distinct phase has a unique combination of chemical formula and modification, defined by its (*a*) prototype (also called structure type), (*b*) Pearson symbol, and (*c*) space group number. One may consider the following example of the *entries* and *distinct phases*. There are the following distinct phases for the well-known titanium dioxide: $TiO_2$ rutile with the space group 136 ($TiO_2$, *tP*6, 136), $TiO_2$ anatase with the space group 141 ($TiO_2$, *tI*12, 141), and $TiO_2$ brookite with the space group 61 ($TiO_2$, *oP*24, 61). Then, the entries for the crystalline structures and physical properties of the titanium dioxide must refer to either of the mentioned distinct phases. Additionally, the phase diagrams must ideally refer to all these distinct phases simultaneously.

The vast majority of the mentioned 115'000 distinct phases have at least one associated crystalline structure. Approximately one seventh of the distinct phases are connected between each other via the phase diagrams. About one tenth of the distinct phases have at least one physical property reported. However, the majority of the distinct phases at the PAULING FILE do not have any of the 8 mentioned physical properties reported. That is why, for their prediction, we decided to apply machine learning, namely the decision-tree regression, drawing the advanced extrapolation from the well known to less known materials. Here we present a proof of concept: how a relatively unsophisticated statistical model trained on the large PAULING FILE dataset predicts a set of the 8 mentioned physical properties from the only crystalline structure, on an example of the nearly 115'000 inorganic compounds, *i.e.* distinct phases.

The materials predictions powered by machine learning have gained traction in the last years [10-11]. Among others, Isayev et al. [11] were training the set of regression and classification models, based on the decision trees, using the ICSD experimental database and the *ab initio* simulation repository. Our present model is similar, although it operates simpler descriptors. Also, according to our comparison, it provides higher quality predictions. This is due to the combination of the higher quality, and the greater size of the training set (*cf.* quality and quantity of the input).

## Results and discussion

**Overview.** The nearly 115'000 (exactly 114'872) distinct phases, missing at least one of 8 selected physical properties, were identified in the current data release of the PAULING FILE. From them, 109'446 (95%) missed all of the 8 physical properties being predicted, and only 865 (0.01%) missed 4 or less physical properties being predicted. Only 14 distinct phases have all 8 physical properties reported, and thus were excluded. Additionally, about 20'000 distinct phases have either incomplete or absent crystalline structures in the PAULING FILE, and thus were also skipped. By the number of the constituent elements, there were 1'461 unary, 14'515 binary, 45'577 ternary, 34'392 quaternary, and 16'296 quinternary suitable distinct phases. About 2'500 remaining phases were of higher orders. Here we report only the results for the number of constituent elements (arity) $N = 1–5$.

**Quality assurance.** The prediction quality of the decision-tree regression is acceptable, and, on average, may even compete with the *ab initio* simulation results. The difference is that the simulation normally requires hours or days of computation time, whereas the machine-learning model yields the results in milliseconds, even with the less powerful hardware. Another difference is that the *ab initio* simulations in practice require careful fine-tuning of the method, whereas the chosen method of machine learning is a black box, where almost no initial setup is needed. The disadvantage of the machine-learning model is that, of course, no physical meaning of predictions is implied. The underlying complex physical phenomena, as well as the lack of training data, lead to poor prediction quality. For example, our model failed on predicting the electrical conductivity, as the mean absolute error was too high. Although the size of the training dataset should not be necessarily huge, there is some minimal threshold — about several thousands of samples.

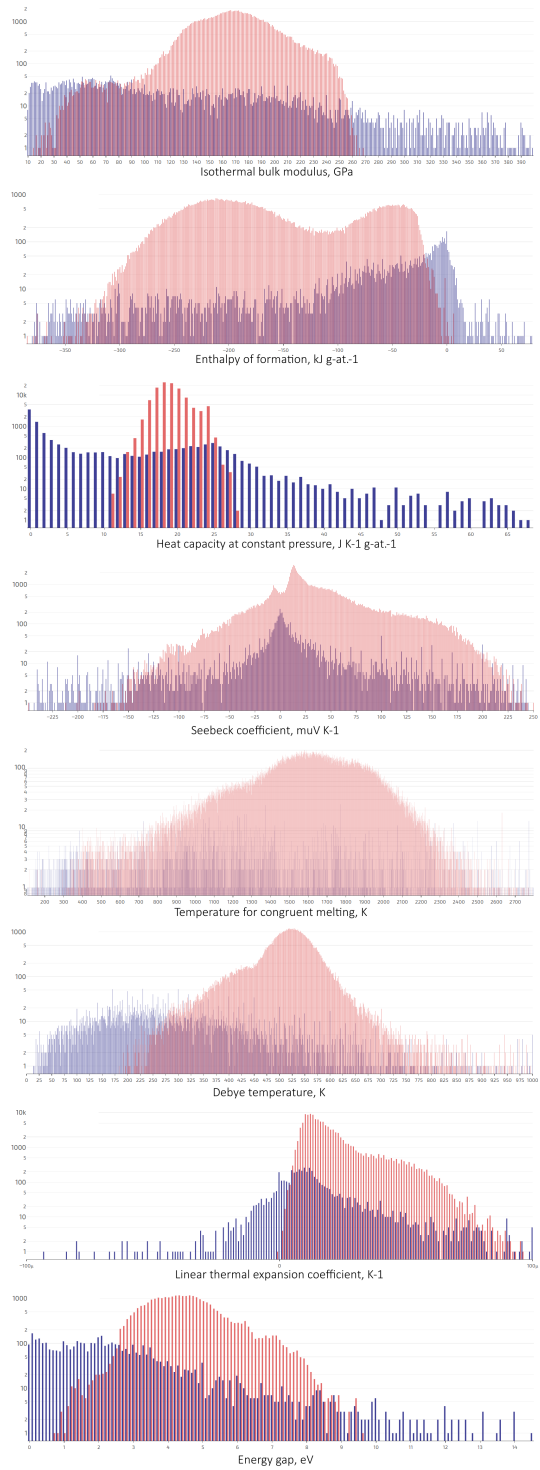

Fig. 1. Distribution of the peer-reviewed (blue) and machine-learning (red) values. From top to bottom: isothermal bulk modulus, enthalpy of formation, heat capacity at constant pressure, Seebeck coefficient, temperature for congruent melting, Debye temperature, linear thermal expansion coefficient, and energy gap (direct or indirect) for insulators.

In Fig. 1 the occurrences of the training and predicted values are compared. The complex distribution shape of predictions suggests that the chosen machine-learning method is very likely to be able to capture the chemical nature of solids. Table 1 presents the quality metrics of the predictions: *mean absolute error* (MAE) and *R-squared coefficient*, a statistical measure of how close the data are to the fitted line. (Best possible R-squared is 1; a constant model always predicting the expected value, disregarding the input, would get an R-squared of 0.) Thus the restricted distribution range in Fig. 1 for the heat capacity at constant pressure, Seebeck coefficient, and linear thermal expansion coefficient should be taken with caution, as they have the lowest R-squared coefficients. Nevertheless although all the R-squared coefficients are relatively low, the trends in predictions are visible quite well (see further). In order to estimate the prediction quality of the binary classifier model "conductor *vs.* insulator", the fraction incorrect (*i.e.* an error percentage) was controlled to be less than 1%.

| Predicted physical property | Units | Mean absolute error | R-squared coeff. |
| --- | --- | --- | --- |
| isothermal bulk modulus | GPa | 45 | 0.45 |
| enthalpy of formation | kJ g-at.$^{-1}$ | 41 | 0.60 |
| heat capacity at constant pressure | J K$^{-1}$ g-at.$^{-1}$ | 4.8 | 0.09 |
| Seebeck coefficient | muV K$^{-1}$ | 89 | 0.12 |
| temperature for congruent melting | K | 294 | 0.63 |
| Debye temperature | K | 95 | 0.33 |
| linear thermal expansion coefficient | K$^{-1}$ 10$^{-5}$ | 1.1 | 0.10 |
| energy gap (if insulator predicted) | eV | 1.1 | 0.30 |

Table 1. The quality estimation of the machine-learning predictions.

Based on the PAULING FILE training dataset, the decision-tree regression and classification models were able to represent 8 physical properties of the wide set of various compounds in an acceptable manner. Nowadays more powerful and accurate machine-learning techniques exist, so the quality of predictions may be further increased. Moreover, combining machine learning with the *ab initio* simulations presents another promising direction.

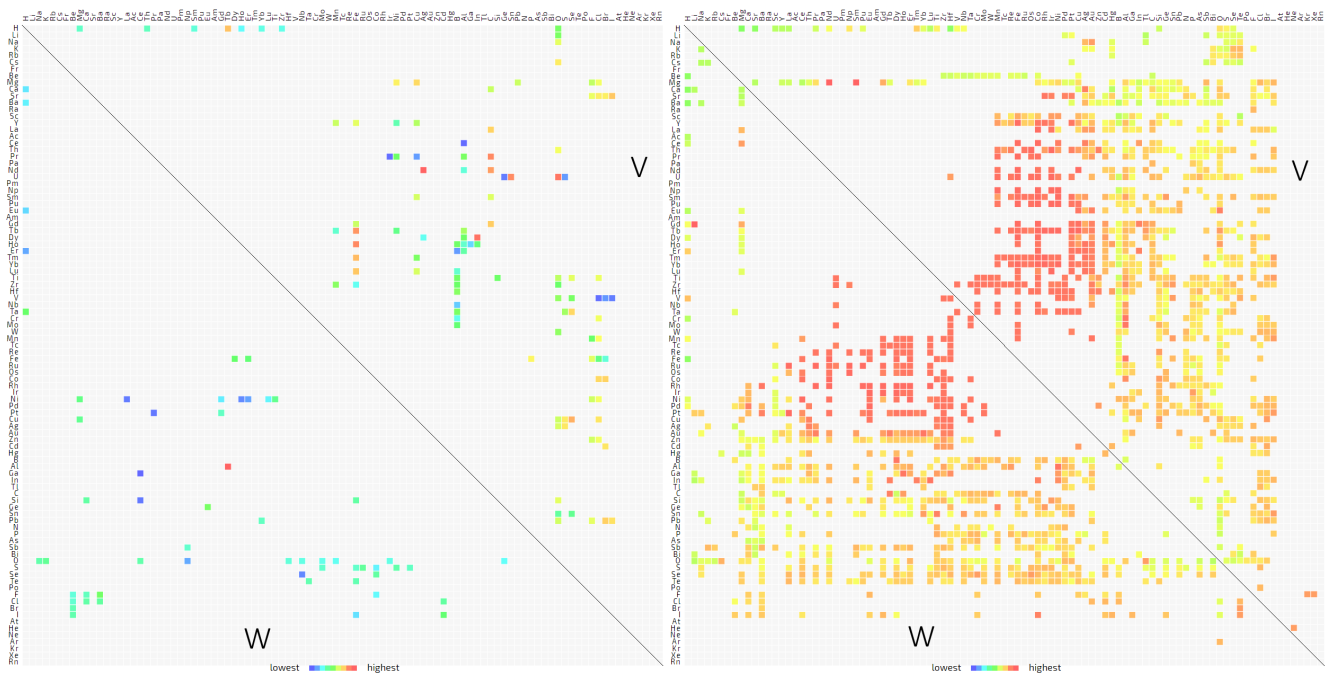

Fig. 2. Periodic numbers $PN_A$ vs. $PN_B$ and the heat capacity at constant pressure (color) for $A_2B$ (denoted by V) and $AB_2$ (denoted by W) compounds, sorted by PNs: peer-reviewed (left) and predicted (right) heatmaps.

**Periodic numbers.** When analyzing the results, we mainly use the periodic numbers of the chemical elements. The periodic number (PN) represents a different enumeration of the elements, emphasizing the role of the valence electrons. In contrast to the atomic number, PN depends on the underlying periodic table of the chemical elements [12]. One can gain new, yet unknown quantitative trends with the aid of heatmaps *e.g.* $PN_A$ vs. $PN_B$ for binary compounds, where the color stands for the particular physical property value, and PN is defined using the Meier's periodic table. For example, in Fig. 2 the heat capacity at constant pressure for $A_2B$ and $AB_2$ compounds (phases) is shown. If more than one peer-reviewed value per distinct phase is available in PAULING FILE, an averaged one is taken. For the predicted values, there is always exactly one value per distinct phase. Interestingly, the maximal heat capacity values were predicted in $A_2B$ phases with the A-elements Mn to Au (PN = 58-76) and the B-elements Ce to Ta (PN = 18-52), as well as in $AB_2$ phases with the A-elements Ce to Ta (PN = 18-52) and the B-elements Cr to Pt (PN = 54-72). The other similar comparisons of the peer-reviewed and predicted data can be plotted online at the Materials Platform for Data Science [9].

**Averaging and dimensionality reduction.** In spite of Fig. 2, we have observed that in about 90% cases within the same chemical element system, the predicted values of the physical properties vary only inconsiderably, not higher than the prediction's MAE. Thus, we averaged the values of the predicted properties within the same chemical element system, which enabled us to deal with the chemical elements only, abstracting from the distinct phases. For instance, rutile, anatase, and brookite, as well as $Ti_2O$, $TiO$ *etc.* were considered simply as Ti-O, or, using PNs, 46-100. To avoid dealing with 6-dimensional spaces – physical property and periodic numbers (or zeros for arity $N < 5$) – we used the principal component analysis (PCA) [13]. This technique aims to reduce the number of dimensions in such an optimal way in order to group similar values together. The PCA returns the linear combinations of the input variables, which are also known as the principal components. This enabled us to visualize all constituent element counts $N = 1$–5 together.

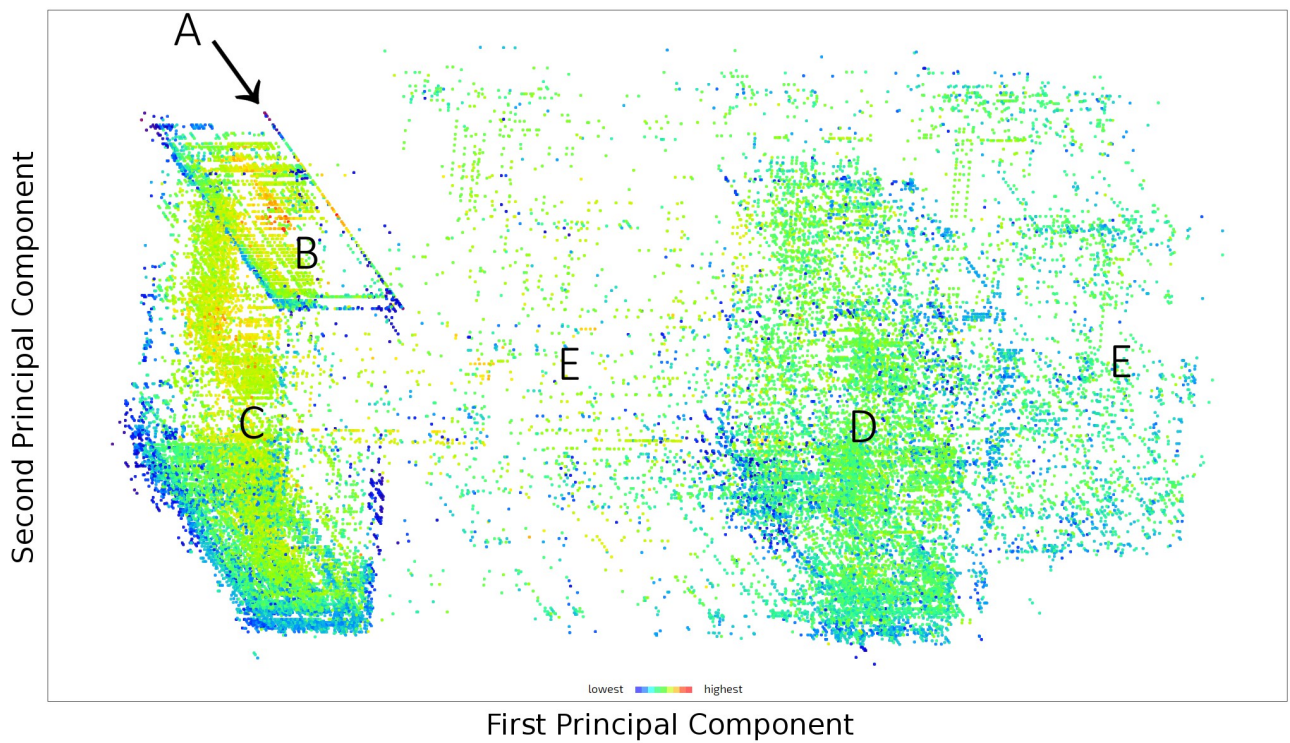

Fig. 3. Principal components of PNs and the temperature for congruent melting (color), arity $N = 1–5$.
Legend: A unaries, B binaries, C ternaries, D quaternaries, E quinternaries.

Being sorted according to their Pauling electronegativities, chemical elements of the considered 114'872 distinct phases were converted to PNs, and the two principal components of PNs were plotted for each of the 8 predicted physical properties, resulting in the heatmaps as shown in Fig. 3 for the temperature for congruent melting. As a result, a surprisingly clear subdivision by the constituent element count was achieved. For example, it is seen that the properties for phases with the arity $N = 1–3$ demonstrate clear patterns (*cf.* Fig. 2), and vague patterns for quaternaries ($N = 4$). On the other hand, the quinternaries ($N = 5$) are not covered enough, and because of that no patterns in quinternaries are seen. The detailed visualizations of patterns for $N \geq 2$ (all distinct phases within the specific chemical element systems) can be found in the supplementary materials or plotted online at the Materials Platform for Data Science [9].

**Neighbor learning.** As a next step, for the 8 considered physical properties we made an attempt to fill in the numerous white spots, as shown in Fig. 3, using an advanced extrapolation technique via the radius-based neighbor learning [14]. All the possible combinations of chemical elements (minus unrealistic ones, such as noble gases, actinides, Tc, Po) were taken to mimic the unknown compounds with the arity $N = 2–5$. The element combinations of the existing PAULING FILE distinct phases were not considered, resulting in nearly seven millions of the totally unknown element combinations. For each element combination, 8 considered physical properties were predicted by the neighbor learning, based on the physical properties, obtained with the decision-tree regression. Thus, each heatmap, as shown in Fig. 3 for the 8 predicted properties, was augmented with the nearly seven million values of the extrapolated property. An example of the enthalpy of formation is shown in Fig. 4. Now the patterns can be seen even for quinternaries ($N = 5$), which present the majority of the extrapolated points.

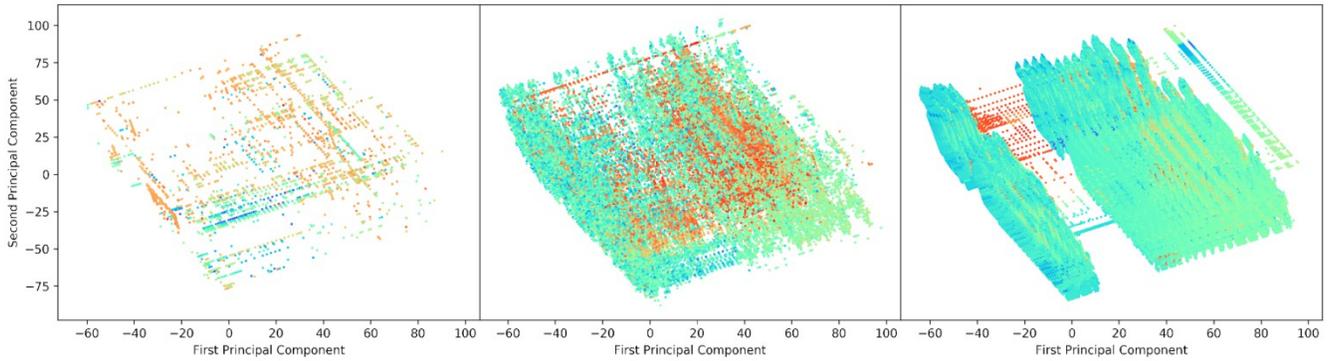

Fig. 4. From left to right: peer-reviewed, decision-tree, and neighbor-learned principal components of PNs and the enthalpy of formation (color). The peer-reviewed and random-forest data stand for the existing PAULING FILE element combinations, whereas the neighbor-learned data stand for about seven millions of the totally unknown element combinations.

**Materials design implementation.** The procedures discussed above were realized in an online application for the materials design [15], which allows predictions of the crystalline structures even for the exotic ranges of properties. This is the reversed task with respect to the previously described decision-tree predictions. As the ranges of properties are wide, the searches are simple and return the existing machine-learning or peer-reviewed data. However as the combination of property ranges becomes nontrivial, no existing data meets the search criteria. Then the design of the new material can be executed, based on the neighbor-learned data (*cf.* Fig. 4, right). The screenshot of the online application is shown in Fig. 5.

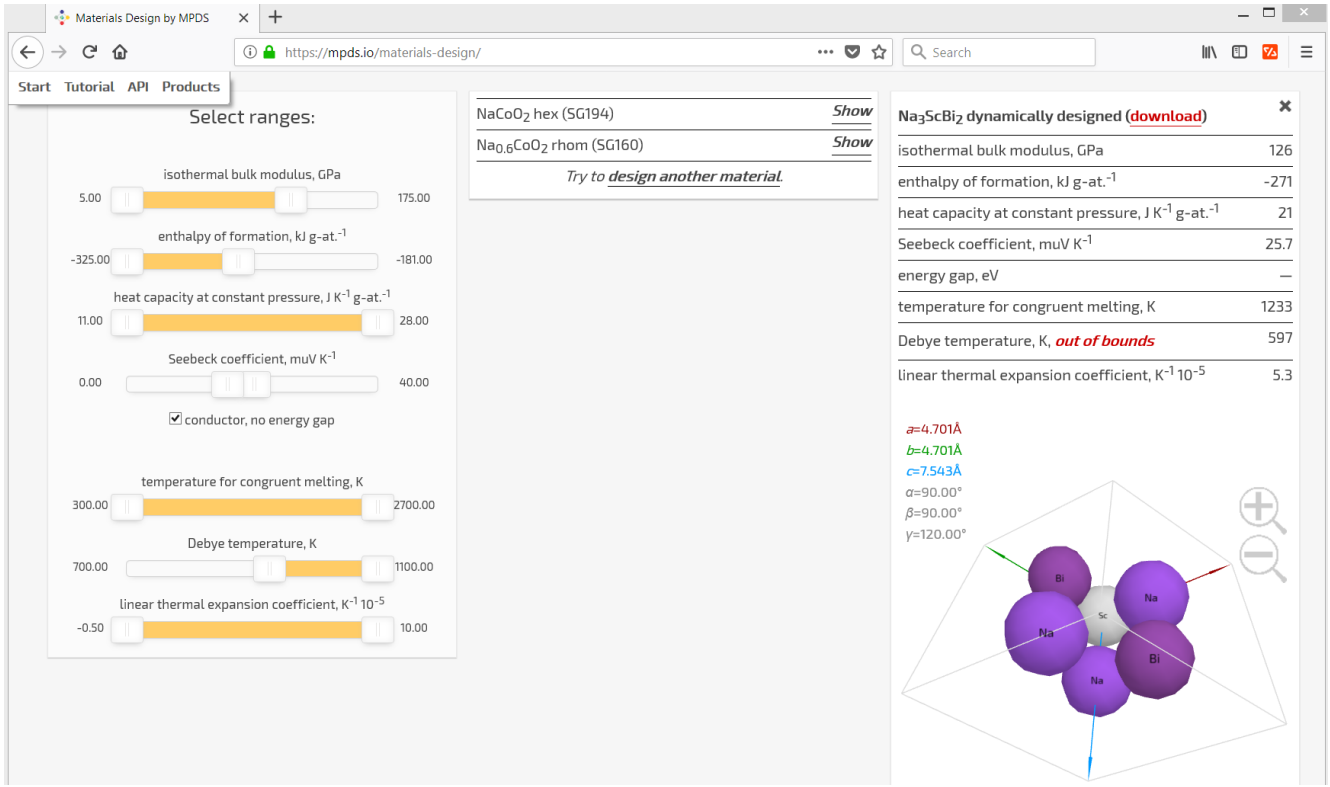

Fig. 5. An online materials design application (https://mpds.io/materials-design). The structure at the right is completely hypothetical and was generated on the fly according to the selected ranges.

## Methods

**Physical property taxonomy.** The physical properties of the PAULING FILE include the experimental data, and to a limited extent, the simulated data of inorganic compounds in the solid, crystalline state. The properties belong to one of the following seven domains: (*a*) electronic and electrical properties, (*b*) magnetic properties, (*c*) mechanical properties, (*d*) optical properties, (*e*) phase transitions, (*f*) superconductivity, and (*g*) thermal and thermodynamic properties. The taxonomy consists of three levels: the mentioned general domains, subdomains, and the particular physical properties. For instance, the domain "electronic and electrical properties" contains the sub-domain "electron energy band structure", which in turn contains the "Fermi energy" property *etc*. Currently there are about 100 sub-domains, and nearly 500 particular numeric physical properties. The 8 particular physical properties considered in this work belong to the most common physical properties extracted by the PAULING FILE project from the world literature. The entire taxonomy was compiled by Prof. Fritz Hulliger (Swiss Federal Institute of Technology in Zurich, Switzerland), Prof. Roman Gladyshevskii (Ivan Franko National University of Lviv, Ukraine) and Dr. Karin Cenzual (University of Geneva, Switzerland). To a certain degree, it reflects the development of the solid state physics during the last century.

**MPDS API.** The MPDS (Materials Platform for Data Science) [9] provides all the PAULING FILE data online via two interfaces: browser-based graphical (GUI) and application programming (API). The GUI is intended for the traditional research, whereas the API serves various software integrations and data-mining scenarios, as described below. Thanks to the API, the materials scientists get unprecedented flexibility and power in an automated analysis of the PAULING FILE data, which is unthinkable within the GUI. In a wider sense, the APIs regulate communications between any kind of the software, be it a chain of the data-mining programs, a simulation platform, or any other big data consumer. The main idea of the API is that all the retrieval functionalities are collected at a single spot and are publicly exposed online. The online APIs normally adhere to the principle of representational state transfer (REST) [16]. The REST presents guiding constraints for client-server software architecture and could be called as the meta-API. The REST is used for the MPDS API, which presents all the PAULING FILE data in a developer-friendly, machine-readable way, using the opened formats, such as CIF and JSON. The MPDS API is carefully documented online.

**Machine-learning regression and classification.** The decision-tree regression and classification is a reliable and simple to use predictive technique. A decision tree is a statistical model, which describes the data going from the observations about some item (*e.g.* a crystalline structure in this work) to the conclusions about the item's target value (*e.g.* the corresponding 8 physical properties). The PAULING FILE data contain crystalline structures linked with the physical properties via the distinct phases. Therefore, it is feasible to train a model on the existing pairs structure-property and then to predict the absent values of properties, based only on the available structures. Multiple decision trees are built by repeatedly resampling training data with replacement, and voting for the trees providing the better prediction accuracy. Such an algorithm is known as a *random forest* [19]. Its presently used state-of-the-art open-source implementation *scikit-learn* (version 0.19.1) takes seconds to train a model from the PAULING FILE data on an average desktop PC [20, 21]. Based on the crystalline structure, the random forest regressor yields all the considered physical properties. Notably, the band gap predictions required additional steps. First, a subdivision of crystalline structures into conductors and insulators was performed via the random forest classifier. This classifier was trained on the particularly imbalanced categorical data, as the PAULING FILE contains nearly five times more conductors than insulators. The random over-sampling (*i.e.* repeating of the under-represented class samples) was therefore applied. Finally, for insulators the particular values of the band gap were obtained.

**Crystalline structure entries.** Currently the PAULING FILE contains about 400'000 crystalline structure entries. The minimal requirement for an entry is a complete set of published cell parameters, assigned to a compound of well-defined composition. Whenever the published data are available, the crystallographic data also include atom coordinates, displacement parameters and experimental diffraction lines, and are accompanied by information concerning preparation, experimental conditions, characteristics of the sample, phase transitions, dependencies of the cell parameters on temperature, pressure, and composition. In order to give an approximate idea of the actual structure, a complete set of atomic coordinates and site occupancies is proposed for an entry, where only a prototype could be assigned by the authors or editor (but no atomic coordinates have been given in the publication). Thus for cell parameters without published atomic coordinates, an entry is prepared for each chemical system and crystalline structure. The crystallographic data are stored as published, but also have been standardized according to the method proposed by Parthé and Gelato [17, 18]. Derived data include atomic environments of the individual atomic sites, and the reduced Niggli cell. The entries are cross-checked both by the editors and editor-in-chief for the inconsistencies within the single entry and comparing the different entries. For 5% of the entries, one or more misprints in the published crystallographic data are detected and corrected. Warnings concerning short interatomic distances, deviations from the nominal composition *etc.* are added in remarks. SI units are forced everywhere, and the crystallographic terms follow the recommendations by the International Union of Crystallography. All the data are extracted from the primary literature. When available, supplementary materials deposited as CIF files or in the other formats are used as sources of data. Crystallographic data, simulated by the *ab initio* calculations or optimized by the other methods, are only considered being confirmed by experimental observations. Distinct entries are created for all the complete refinements reported in a particular paper. For example, for a continuous solid solution between two ternary compounds, there will be three entries: one for each boundary ternary composition and one for the quaternary system. The latter may contain a remark describing the composition dependence of the cell parameters. The preference is given to values determined under ambient conditions.

**Crystalline structure descriptors and training.** The peer-reviewed crystalline structures and their corresponding 8 physical properties were obtained via the MPDS API. The aim was to train the machine-learning model on all the available data, so that the relation between the crystalline structures and the physical property values could be found in a purely statistical manner. Any crystalline structure was populated to a certain relatively large fixed volume of a minimum one cubic nanometer. Then the crystalline descriptor was constructed using the PNs of atoms and the lengths of their radius-vectors. (Here the term descriptor stands for the compact information-rich representation, allowing the convenient mathematical treatment of the encoded complex data, *i.e.* crystalline structure.) To evaluate the prediction, random 33% crystalline structures together with the corresponding physical properties were isolated and excluded from training. Next, the trained machine-learning model had to predict these physical properties based on the isolated crystalline structures. The factual and predicted values were compared, and the quality metrics were calculated. The evaluation process was repeated 30 times to achieve a statistical reliability. Here a bibliographic issue arose, as the several physical property values or several crystalline structures could be reported for a distinct phase. In those cases the averaged values were considered. Finally, the evaluated model was trained on all the data.

**Crystalline structure disorder.** About 55% of all the PAULING FILE crystalline structures are disordered. The disorder was solved by structure ordering; however only a very small number of ordered structures was randomly considered. It was shown, that sampling only 6 ordered structures (instead of possible millions) already leads to a relatively stable averaged prediction. That means that the machine-learning method is not very sensitive to the atomic permutations. Of course, increasing the number of the ordered samples yields more stable and consistent predictions. Here however the speed was the limiting factor, as processing each sample requires some computational resources.

**Neighbor learning.** A radius-based neighbor learning as implemented in the *scikit-learn* program toolkit (version 0.19.1) was used. An exceptional quality of extrapolations was demonstrated. The values for 8 physical properties found by the neighbor learning from the neighbor chemical elements and the values predicted by decision-tree regression from the crystalline structure differ on average by only 5-10%.

**Materials design.** The reversed task of predicting the crystalline structures by the given combination of properties (Fig. 5) was done as follows. With the extrapolation of the 8 physical properties via the neighbor learning, the number of the new chemical element combinations was increased in about ten times. Thus, the probability of finding at least some chemical elements for an arbitrary combination of the input property ranges was also increased considerably. The obtained chemical elements in a previously unknown combination are searched in the MPDS via the fuzzy chemical matching, *i.e.* by the similar PNs. Then the elements in the found chemically similar structures are replaced with the sought elements. Under the naive assumption, that the cell and the inter-atomic distances remain the same, this technique works surprisingly well. Finally, the random forest regression is executed; the results are scored and returned.

**Supplementary**

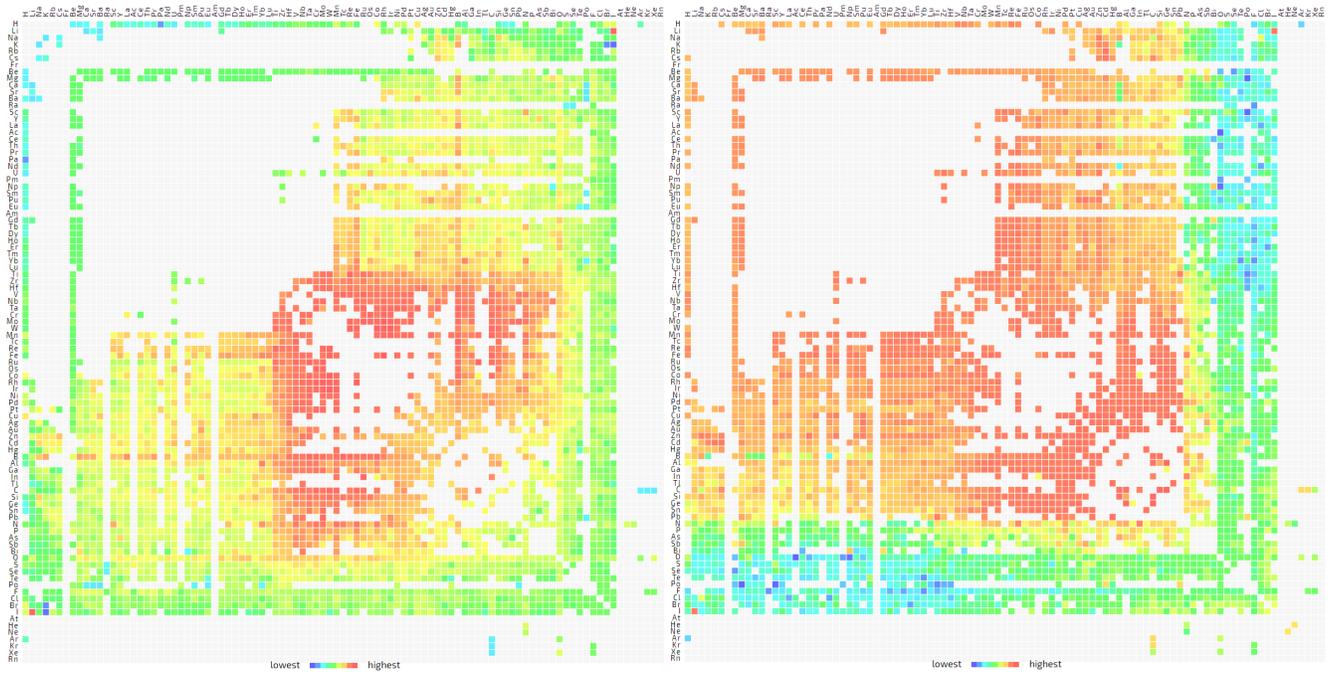

Fig. A. Heatmaps of $PN_A$ *vs.* $PN_B$ *vs.* predicted values (colors) for all distinct binary phases within the specific chemical element systems. Left: isothermal bulk modulus. Right: enthalpy of formation.

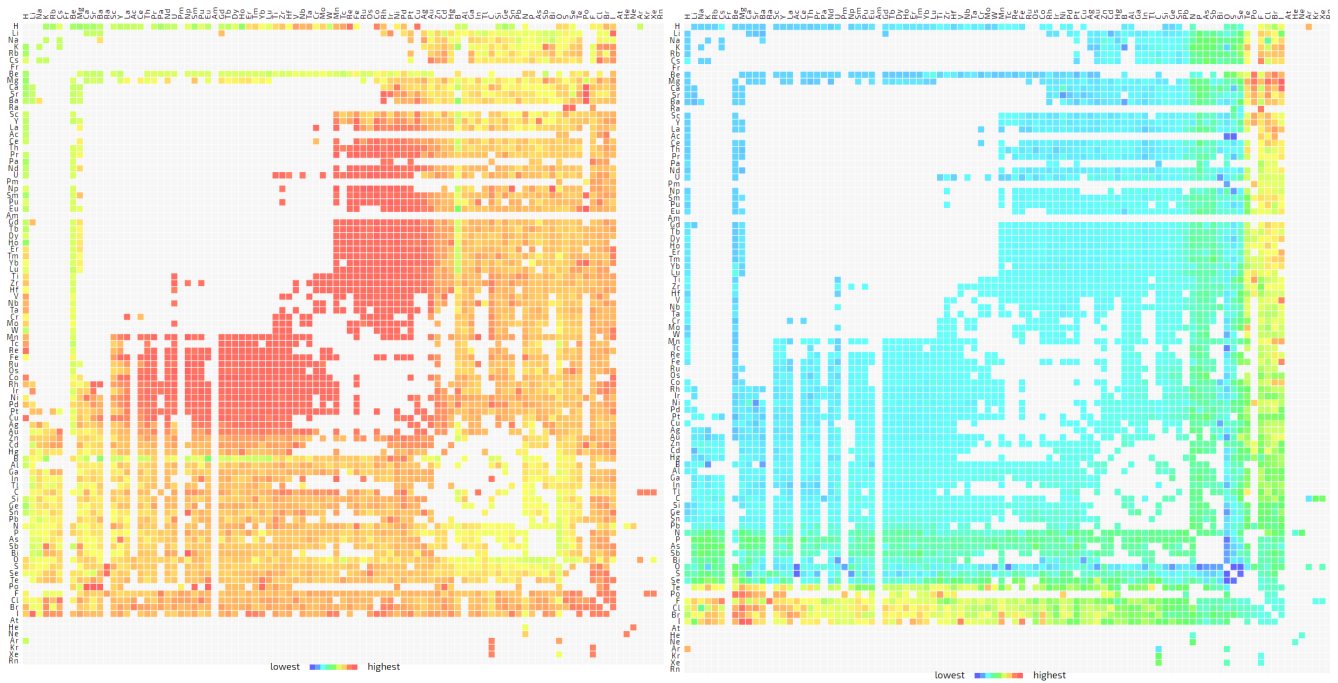

Fig. B. Heatmaps of $PN_A$ *vs.* $PN_B$ *vs.* predicted values (colors) for all distinct binary phases within the specific chemical element systems. Left: heat capacity at constant pressure. Right: Seebeck coefficient.

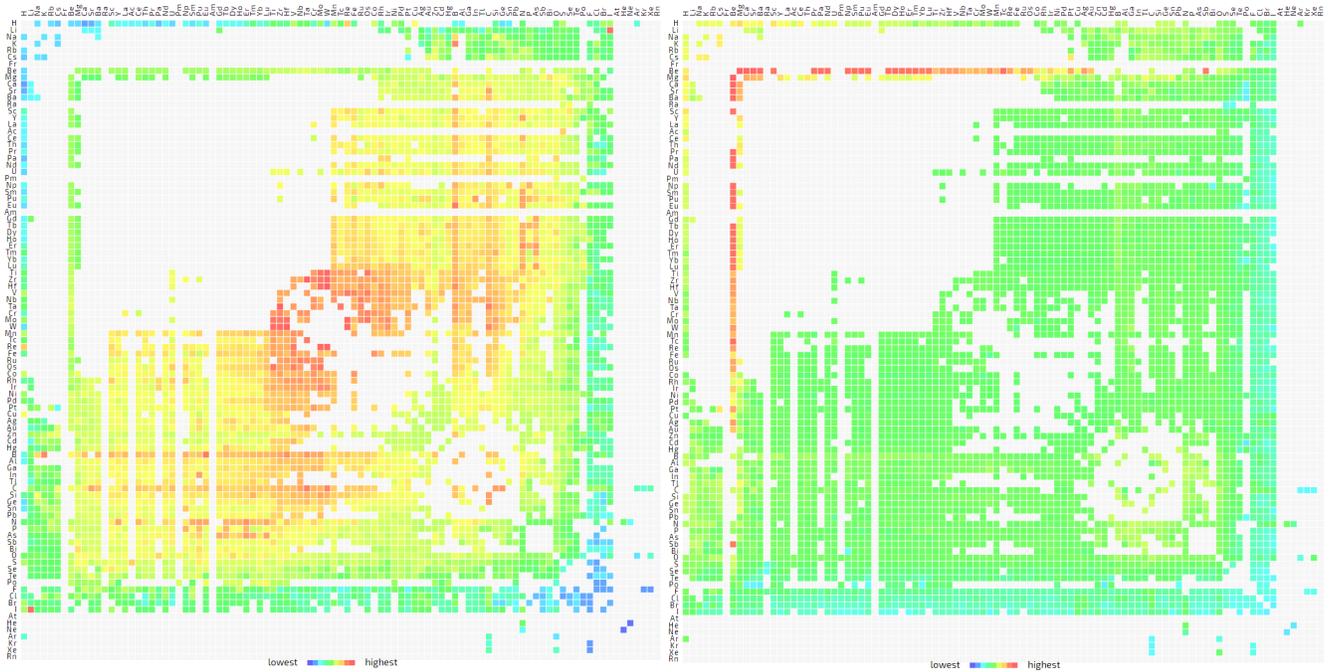

Fig. C. Heatmaps of $PN_A$ vs. $PN_B$ vs. predicted values (colors) for all distinct binary phases within the specific chemical element systems. Left: temperature for congruent melting. Right: Debye temperature.

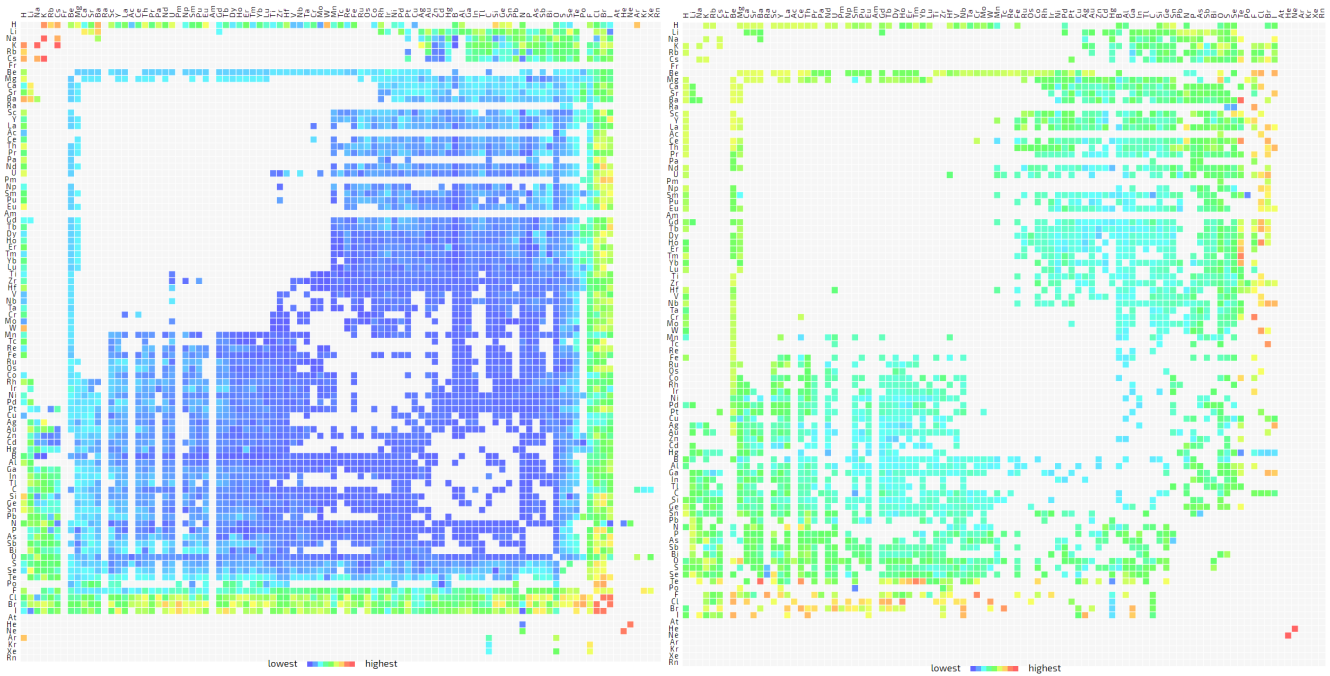

Fig. D. Heatmaps of $PN_A$ vs. $PN_B$ vs. predicted values (colors) for all distinct binary phases within the specific chemical element systems. Left: linear thermal expansion coefficient. Right: energy gap (direct or indirect) for insulators.